\documentclass[12pt]{article}
\usepackage[mathscr]{eucal}
\usepackage{amsmath,amssymb,amscd}
\usepackage{color}
\usepackage{epsfig}

\def \beq{ \begin{equation} }
\def \eeq{\end{equation}}

\begin{document}
\begin{center}
\begin{huge}
\textbf{On polygonal relative equilibria\\
in the \(N\)-vortex problem\\\( \)}\\
\end{huge}
\begin{large}
\textbf{M. Celli, E.A. Lacomba and E. P\'erez-Chavela\\\( \)}\\
\end{large}
Departamento de Matem\'aticas\\
Universidad Aut\'onoma Metropolitana-Iztapalapa\\
Av. San Rafael Atlixco 186, Col. Vicentina, M\'exico, D.F. 09340, M\'exico
\end{center}

\vskip1cm

\begin{abstract}
Helmholtz's equations provide the motion of a system of \(N\)
vortices which describes a planar incompressible fluid with zero
viscosity. A relative equilibrium is a particular solution of
these equations for which the distances between the vortices are
invariant during the motion. In this article, we are interested in
relative equilibria formed of concentric regular polygons of
vortices. We show that in the case of one regular polygon
(and a possible vortex at the center)
with more than three vertices
(two if there is a vortex at the center),
a relative equilibrium requires equal vorticities
(on the polygon).
We also determine all the relative equilibria with two concentric
regular \(n\)-gons and the same vorticity on each \(n\)-gon. This
result completes the classical studies for two regular \(n\)-gons when
all the vortices have the same vorticity or when the total
vorticity vanishes.
\end{abstract}

\vskip1cm

\renewcommand{\thesection}{\Roman{section}}
\section{Introduction}
\renewcommand{\thesection}{\arabic{section}}

In this paper, we are interested in polygonal relative equilibria
of \(N\) vortices in a planar incompressible fluid with zero
viscosity. Their motion is given by Helmholtz's equations
(Helmholtz, 1858):
\[\dot{z}_k=i\sum _{l\ne k} \Gamma _l\frac{z_k-z_l}{|z_k-z_l|^2}
=i\sum _{l\ne k} \frac{\Gamma _l}{\bar{z}_k-\bar{z}_l},\] where
\(\Gamma _1\), \(\ldots \), \(\Gamma _N\) are the vorticities,
\(z_1\), \(\ldots \), \(z_N\) are the positions, seen as complex
numbers, and \(i=\sqrt{-1}\). These equations are integrable
only for \(N=2\) or \(3\).\\

An \(N\)-vortex motion is said to be a {\it relative equilibrium}
when the mutual distances between the vortices are constant. This
is equivalent to having, at a given time:
\[\dot{z}_l-\dot{z}_k=i\omega (z_l-z_k),\]
where \(\omega \) is the angular velocity. For \(\omega =0\), all
the velocities are equal. The motion is either an absolute
equilibrium or a rigid translation with constant non-zero
velocity. For \(\omega \ne 0\), the motion is a rotation with
constant angular velocity \(\omega \) around a fixed center
\(\Omega \), such that:
\[\dot{z}_k=i\omega (z_k-\Omega )\cdot \]
When the total vorticity is different from \(0\),
the motion is an absolute equilibrium or a rotation with center
the {\it center of vorticity} defined as:
\[\frac{1}{\sum _{k=1} ^N \Gamma _k}\sum _{k=1} ^N \Gamma _kz_k\cdot \]
The determination of relative equilibria is a difficult problem.
It is only since 2009 that we have known that for a given value of
the four vorticities, there is a finite number of four-vortex
relative equilibria, up to similarities (Hampton and Moeckel,
2009). Moreover, the proof of this result requires sophisticated
mathematical tools (such as the BKK theory of sparse polynomial
systems) and the use of a computer. A good introduction to the
problems and methods of vortex dynamics can be found in Newton
(2001) and in Aref (2007). A more specific introduction to
relative equilibria in the \(N\)-vortex problem can be found in
Aref et al. (2002).\\

It is easy to check that when the \(N\) vortices have the same
vorticity and are located at the vertices of a regular polygon,
they form a relative equilibrium. The motion is a rotation around
the center of the polygon, with non-zero angular velocity.
Adding a vortex at the center, we obtain
the same kind of motion, or an absolute equilibrium.
We sometimes have relative equilibria in the case of
several regular concentric or nested \(n\)-gons, with the same
vorticity on each polygon, and a possible vorticity at the
center.\\

Phenomena related to various fields of science motivate the study
of such motions. Polygonal \(N\)-vortex relative equilibria were
detected in superfluid Helium (Yarmchuk et al., 1979). The great
advance in satellite observations in the last years has allowed
precise stu\-dies of various atmospherical phenomena, among them are
hurricanes. In many of them, the eyewall (which is the ring
surrounding the eye of the hurricane, where the wind and rain are
the strongest) has a polygonal shape. Some of these eyewalls were
first discovered numerically, and then observed by satellite. A
remarkable example is Hurricane Isabel (Kossin and Schubert, 2001;
Kossin and Schubert, 2004), which showed a regular pentagonal
pattern, with a vortex at the center. Later, other patterns were
observed in this hurricane. In one of them, eight small vortices
could be seen, creating after some time a square of vortices of
larger size. Concentric eyewalls have been observed in some
intense hurricanes, a nice example is the triple eyewall of
Hurricane Juliette (McNoldy, 2004). This phenomenon happens when
outer rainbands manage to organize into a new eyewall. This outer
eyewall moves inward and strengthens thanks to the moisture and
angular momentum of the inner eyewall, whereas the inner eyewall
dissipates. Eventually the outer eyewall completely replaces the
inner one. The presence of polygonal patterns in hurricanes and
its consequences are still very enigmatic and raise important and
difficult questions.\\

In this article, we are particularly interested in relative
equilibria formed of one or two concentric regular polygons.
In Section 1, we show that, in the case of one regular polygon
(with a possible vortex at its center)
formed of \(N\ge 4\) vortices
(\(N\ge 3\) if there is a vortex at the center),
the motion is a relative
equilibrium only for equal vorticities (on the polygon).
The relative equilibria
formed of two concentric \(n\)-gons, with the same vorticity on each
\(n\)-gon, are known for some values of the vorticities (Havelock,
1931; Aref et al., 2002). In Section 2, we provide an exhaustive
classification of all these relative equilibria, which is valid
for every value of the vorticity on each polygon.

\vskip1cm

\renewcommand{\thesection}{\Roman{section}}
\section{One polygon with a possible vortex at its center}
\renewcommand{\thesection}{\arabic{section}}

In the three-vortex problem, it is easy to check that when the
vortices are at the vertices of an equilateral triangle, they form
a relative equilibrium for every value of the vorticities. When
the total vorticity does not vanish, the motion is a rotation
around the center of vorticity with constant non-zero angular
velocity. When the total vorticity vanishes, the motion is a
translation with constant non-zero velocity. The aim of this
section is to show that, on the other hand, a regular polygon
formed of \(N\ge 4\) vortices is in relative equilibrium only
for equal vorticities. Moreover, a configuration formed of a polygon
with \(N\ge 3\) vortices and a vortex at its center is in
relative equilibrium only for equal vorticities on the polygon.\\

Our result provides a possible explanation for a strange
phenomenon observed in some numerical si\-mu\-la\-tions of
hurricanes. In various Weather Research and Forecasting
si\-mu\-la\-tions (see Davis et al.,
2008, for the case of Hurricane Katrina), many triangular eyewalls
could be observed, whereas all the vortex polygons physically
observed in the real eyewall of the same hurricanes do have at
least four vertices. In fact, our result highlights a stability
property (with regard to the vorticity parameter) peculiar to
vortex triangles. This could explain a preferential convergence of
some numerical algorithms to these more stable con\-fi\-gu\-rations.\\

The arguments of our demonstration were inspired by the proof of a
formally similar result in celestial mechanics. In that problem,
we have to consider relative equilibria of \(N\) punctual bodies
which interact through gravitation, with masses \(m_1\), \(\ldots
\), \(m_N\), and positions \(\vec{r}_1\), \(\ldots \),
\(\vec{r}_N\). The motion of the bodies is given by Newton's
equations:
\[\ddot{\vec{r}}_k=\sum _{l\ne k}
m_l\frac{\vec{r}_l-\vec{r}_k}{||\vec{r}_l-\vec{r}_k||^3} \cdot \]
In Perko and Walter (1985) and Elmabsout (1988), it is proved
that, for a polygonal relative equilibrium with \(N\ge 4\)
celestial bodies, all the masses have to be equal. Although
Newton's equations describe a physical system totally different
from an incompressible fluid, they are formally close to
Helmholtz's equations. That is why some results and methods
relating to Newton's equations can be adapted to the study of
Helmholtz's equations, and conversely. However, the existence of
negative vorticities (whereas a mass is always positive) and
systems with total vorticity zero makes the \(N\)-vortex problem
more difficult in a certain sense. Fortunately, the exponent \(2\)
instead of \(3\) in the denominator of the equations also makes
the \(N\)-vortex problem easier in another sense.\\

The idea is to make use of the properties of two virtual velocity
fields, that we are going to superimpose to the velocity field we
study. Then we will show that our new velocity field satisfies
simple equations, which will happen to have exactly one solution,
that we will define as our third virtual velocity field. These
three virtual velocity fields will be generated by sources, sinks
and vortices at the vertices \(1\), \(\rho =e^{i2\pi /N}\), \(\rho
^2\), \(\ldots \), \(\rho ^{N-1}\) of a regular polygon with
center the origin and radius \(1\). We will agree that a source
with intensity \(\Gamma >0\) at point \(z_0\) generates the
following velocity field:
\[v(z)=\Gamma \frac{z-z_0}{|z-z_0|^2}\cdot \]
We will see a sink with intensity \(\Gamma >0\) as a source with
negative intensity \(-\Gamma \), and a vortex with vorticity
\(\Gamma \) as a source with imaginary intensity \(i\Gamma \). The
superposition, at the same point, of a source (or a sink) and a
vortex can be seen as a source with intensity an imaginary number
\(a+ib\). Unlike the velocity fields generated by pure vortices,
the fields generated by complex sources
are not necessarily incompressible.\\

Such complex sources can exist in nature. However, according to
Euler's equations, their intensity is constant only when they are
pure imaginary numbers, which corresponds to the case of vortices.
Nevertheless, in this section, we are going to consider systems
with complex sources with constant intensity, whose motions are
solutions of generalized Helmholtz's equations with complex
intensities, where the terms \(i\Gamma _k\) of the classical
equations with vortices were changed for complex intensities
\(a+ib\). However, we will have to keep in mind that here, these
motions with complex sources are only {\it auxiliary variables},
algebraically close to the vortex motions that they generalize.
But they will help us in finding simple conditions satisfied by
real systems of vortices. As in the case of vortices, the motions
of complex sources can be or not be relative equilibria, with a
fixed center of rotation in the case \(\omega \ne 0\). One can
check that the equations of a relative equilibrium are the same
as the equations of the previous section for the case of vortices.\\

Let us define our virtual velocity field \(1\) as the field
generated by \(N\) sources with complex intensities \(1\), \(\rho
^{-1}\), \(\rho ^{-2}\), \(\ldots \), \(\rho ^{-(N-1)}\) at the
vertices \(1\), \(\rho \), \(\rho ^2\), \(\ldots \), \(\rho
^{N-1}\) of a regular polygon. Thanks to symmetry
considerations, we can easily see that this configuration is
rigidly translating with non-zero velocity. As a matter of fact,
the velocity \(v_0\) of the source located at point \(1\) is the
sum of the contributions of the sources located at \(\rho \),
\(\rho ^2\), \(\ldots \), \(\rho ^{N-1}\), with intensities \(\rho
^{-1}\), \(\rho ^{-2}\), \(\ldots \), \(\rho ^{-(N-1)}\). And the
velocity \(v_1\) of the source located at point \(\rho \) is the
sum of the contributions of the sources located at \(\rho ^2\),
\(\rho ^3\), \(\ldots \), \(\rho ^{N-1}\), \(\rho ^N=1\), with
intensities \(\rho ^{-2}\), \(\rho ^{-3}\), \(\ldots \), \(\rho
^{-(N-1)}\), \(\rho ^{-N}=1\). From \(v_0\) to \(v_1\), the
positions of the sources are multiplied by \(\rho \), whose
modulus is \(1\), and the intensities are divided by \(\rho \). So
the two velocities have to be equal, and equal to the velocities
of the other sources.
A simple computation shows that these velocities do not vanish.\\

Let us define our virtual velocity field \(2\) as the field
generated by \(N\) sources whose intensities are all equal to
\(i\) (or by \(N\) vortices whose vorticities are all equal to
\(1\)) at the vertices \(1\), \(\rho \), \(\rho ^2\), \(\ldots \),
\(\rho ^{N-1}\) of the regular polygon. This configuration is a
relative equilibrium
with non-zero angular velocity and center \(0\).\\

Let us define our virtual velocity field \(3\) as the field
generated by \(N\) sources with intensities \(1\), \(-e^{-i\pi
/N}\), \(e^{-i2\pi /N}\), \(\ldots \), \(e^{-i(N-1)\pi /N}\) at
the vertices \(1\), \(\rho \), \(\rho ^2\), \(\ldots \), \(\rho
^{N-1}\) of the regular polygon. This definition is only relevant
when \(N\) is odd, as the first and \(N\)-th vorticities have to
be the same: \((-1)^Ne^{-iN\pi /N}=1\). Thanks to symmetry
considerations, we can see that this configuration is an absolute
equilibrium. As a matter of fact, the velocity of the source
located at point \(1\), for instance, is the sum of the
contributions of the sources located at \(\rho \) and \(\rho
^{N-1}\), \(\rho ^2\) and \(\rho ^{N-2}\), \(\ldots \), \(\rho
^{(N-1)/2}\) and \(\rho ^{(N+1)/2}\). A simple computation shows
that the contribution of the source located at \(\rho \) is the
opposite of the contribution of the source located at \(\rho
^{N-1}\). The same results holds for the contributions of the
sources located at \(\rho ^2\) and \(\rho ^{N-2}\), \(\ldots \),
\(\rho ^{(N-1)/2}\) and \(\rho ^{(N+1)/2}\). Thus, the source
located at \(1\) has zero velocity.
This argument can also be applied to the other sources,
so all of them have zero velocity.\\

\begin{figure}[h]
  \centering
     \includegraphics[width=7cm]{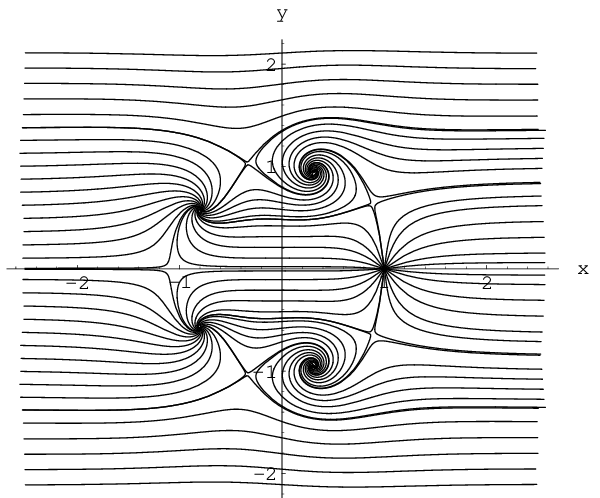}
     \includegraphics[width=7cm]{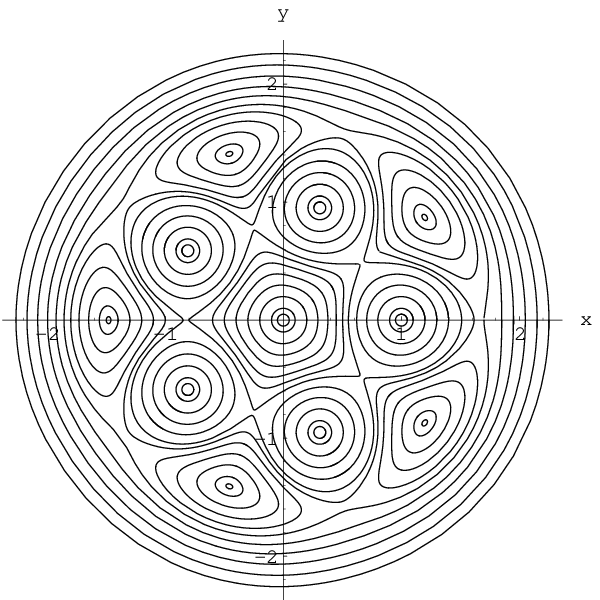}
     \includegraphics[width=7cm]{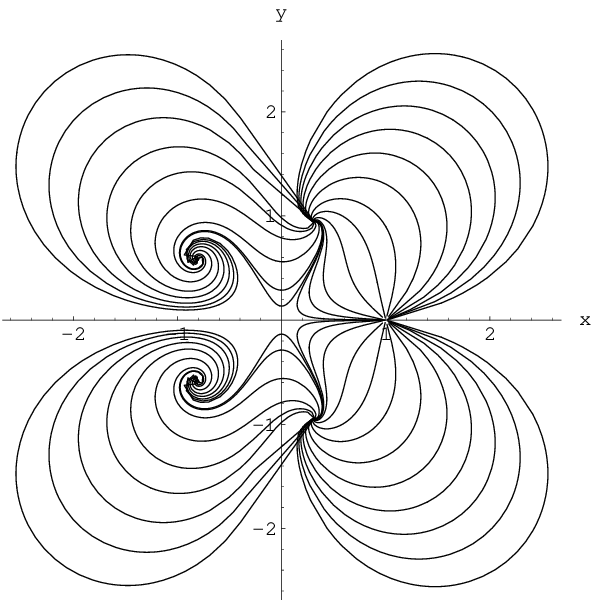}
       \caption{Pathlines of virtual fields \(1\) (at the top, on the left),
       \(2\) (at the top, on the right)
       and \(3\) (at the bottom), in the case \(N=5\).}
       \label{fields}
       \end{figure}

In figure \ref{fields}, we plotted the pathlines for these three
virtual velocity fields in the rotating frame where the polygon
does not move, in the case \(N=5\). We can see various sets of
closed pathlines of virtual field \(2\): around the five vortices
(with infinite velocity), and around the origin and the vertices
of another pentagon (with zero velocity, so they are equilibrium centers).\\

Let us consider the velocity field generated by complex sources
\(\Gamma _0\), \(\ldots \), \(\Gamma _{N-1}\) in relative equilibrium
at the vertices \(1\), \(\rho \), \(\rho ^2\), \(\ldots \),
\(\rho ^{N-1}\) of the regular polygon.
Let us first assume that the angular velocity
\(\omega \) does not vanish. What can we obtain after superimposing
this velocity field and \(\lambda \) times (where \(\lambda \) is
a complex number) rigidly translating virtual field \(1\)? In
other words, what can we obtain after a change of sources of the
form:
\[(\Gamma _0', \ldots , \Gamma _{N-1}')
=(\Gamma _0, \ldots , \Gamma _{N-1})+\lambda (1,\rho ^{-1}, \ldots
,\rho ^{-(N-1)})?\] According to the linearity in the sources of
Helmholtz's equations, the velocity field obtained, which is the
superposition of two relative equilibria, will be in relative
equilibrium with angular velocity \(\omega \) and center \(\Omega
+i\lambda v/\omega \) (\(\Omega \) is the center of the relative
equilibrium that we are considering and \(v\) is the translation
velocity of field \(1\)). So we can choose \(\lambda \) such that
the new center is at the origin. Thus, using a change of
sources, we are now able to restrict our study to
relative equilibria with center at the origin.\\

What can we obtain now after superimposing our new field and \(\mu
\) times (where \(\mu \) is a real number) virtual field \(2\) in
relative equilibrium? In other words, what can we obtain after a
change of sources of the form:
\[(\Gamma _0'', \ldots , \Gamma _{N-1}'')
=(\Gamma _0', \ldots , \Gamma _{N-1}')+\mu (i,\ldots ,i)?\]
According to the linearity in the sources of Helmholtz's
equations, the velocity field obtained, which is the superposition
of two relative equilibria, will be in relative equilibrium, with
cons\-tant polygon center at \(0\) and angular velocity \(\omega
+\mu \omega '\), where \(\omega \) is the angular velocity of the
field that we are considering and \(\omega '\) is the angular
velocity of field \(2\). So we can choose \(\mu \) such that the
new angular velocity vanishes. Thus, using changes of sources,
we are now able to restrict our study to
absolute equilibria.\\

Using a similar change of sources, we can deal with the case
with vanishing angular velocity \(\omega \). Subtracting a
suitable multiple of rigidly translating field \(1\)
(and no multiple of field \(2\)),
as we did in the case \(\omega \ne 0\),
we obtain zero translation velocity,
i.e. an absolute equilibrium again.\\

The absolute equilibria that we are looking for are characterized
by the following equation:
\[\left (
\begin{array}{ccccc}
0 & \frac{1}{1-\rho ^{-1}} & \frac{1}{1-\rho ^{-2}} & \ldots & \frac{1}{1-\rho ^{-(N-1)}}\\
\frac{1}{\rho ^{-1}-1} & 0 & \frac{1}{\rho ^{-1}-\rho ^{-2}} & \ldots & \frac{1}{\rho ^{-1}-\rho ^{-(N-1)}}\\
\vdots & \vdots & \vdots & & \vdots \\
\frac{1}{\rho ^{-(N-1)}-1} & \frac{1}{\rho ^{-(N-1)}-\rho ^{-1}} &
\frac{1}{\rho ^{-(N-1)}-\rho ^{-2}} & \ldots & 0
\end{array}
\right ) \left (
\begin{array}{c}
\Gamma _0''\\
\vdots\\
\Gamma _{N-1}''
\end{array}
\right ) =\left (
\begin{array}{c}
0\\
\vdots\\
0
\end{array}
\right ),\] which is equivalent to: \(M\Gamma =0\), where \(\Gamma
=(\Gamma _0'', \ldots , \Gamma _{N-1}'')\) and:
\[M=\left (
\begin{array}{ccccc}
0 & \frac{1}{1-\rho ^{-1}} & \frac{1}{1-\rho ^{-2}} & \ldots & \frac{1}{1-\rho ^{-(N-1)}}\\
\frac{1}{1-\rho ^{-(N-1)}} & 0 & \frac{1}{1-\rho ^{-1}} & \ldots & \frac{1}{1-\rho ^{-(N-2)}}\\
\vdots & \vdots & \vdots & & \vdots \\
\frac{1}{1-\rho ^{-1}} & \frac{1}{1-\rho ^{-2}} & \frac{1}{1-\rho
^{-3}} & \ldots & 0
\end{array}
\right )\cdot \]

We saw that for \(\Gamma =(1,\rho ^{-1},\ldots ,\rho ^{-(N-1)})\)
(the sources which generate virtual field \(1\)), the
configuration is rigidly translating with non-zero velocity. This
is equivalent to:
\[M(1,\rho ^{-1},\ldots ,\rho ^{-(N-1)})=v(1,\rho ^{-1},\ldots ,\rho ^{-(N-1)}),\]
where \(v\) is the velocity of translation. Thus: \(\Gamma
=(1,\rho ^{-1},\ldots ,\rho ^{-(N-1)})\)
is an eigenvector of \(M\) with non-zero corresponding eigenvalue.\\

We also saw that for \(\Gamma =(i,\ldots ,i)\) (the sources which
generate virtual field \(2\)), the configuration is a relative
equilibrium with non-zero angular velocity. This is equivalent to:
\[M(i,\ldots ,i)=\omega (i,\ldots ,i),\]
where \(\omega \) is the angular velocity. Thus: \(\Gamma
=(i,\ldots ,i)\)
is an eigenvector of \(M\) with non-zero corresponding eigenvalue.\\

Lastly, we saw that for \(\Gamma =(1,-e^{-i\pi /N},e^{-i2\pi /N},
\ldots , e^{-i(N-1)\pi /N})\) (the sources which generate virtual
field \(3\)), the configuration is an absolute equilibrium. Thus,
this vector
is an eigenvector of \(M\) whose corresponding eigenvalue is \(0\).\\

In fact, it is easy to see that the fact that these three systems
of sources are eigenvectors follows from the only ``circulant''
feature of these vectors (multiplying a component by a constant
coefficient, we obtain the next one) and the matrix \(M\)
(shifting all the terms of a line one index to the right, we
obtain the following line). Using this argument, we can see
that, more generally, the \(N\) vectors \((1,\rho ^k,\rho
^{2k},\ldots ,\rho ^{(N-1)k})\), where \(k=0\), \(\ldots \),
\(N-1\), are eigenvectors of \(M\). Besides, we can check that
they form an orthogonal basis. The corresponding eigenvalues have
the following expression:
\[\lambda _k
=\sum _{l=1} ^{N-1} \frac{\rho^{kl}}{1-\rho ^{-l}}
=\frac{1}{2}\sum _{l=1} ^{N-1} \left (\frac{\rho^{kl}}{1-\rho
^{-l}} +\frac{\rho^{-kl}}{1-\rho ^l}\right ) =\frac{1}{2}\sum
_{l=1} ^{N-1} \frac{\rho^{(k+1)l}-\rho^{-kl}}{\rho ^l-1}\]
\[=\frac{1}{2}\sum _{l=1} ^{N-1}
\frac{1}{\rho ^l-1} \sum _{m=-k} ^k (\rho^{(m+1)l}-\rho^{ml})
=\frac{1}{2}\sum _{m=-k} ^k\sum _{l=1} ^{N-1} \rho ^{ml}
=\frac{1}{2}(N-1-2k)\cdot \] If \(N\) is even, there is no
absolute equilibrium. If \(N\) is odd, the values of the sources
which generate virtual field \(3\) (which is the eigenvector
obtained for \(k=(N-1)/2\))
provide the unique absolute equilibrium.\\

Let us now consider a physical polygonal relative equilibrium
formed of \(N\ge 4\) vortices (the sources are pure imaginary
numbers). If \(N\) is even, we have:
\[(\Gamma _0'',\ldots ,\Gamma _{N-1}'')
=(\Gamma _0,\ldots ,\Gamma _{N-1})+\lambda (1,\rho ^{-1},\ldots
,\rho ^{-(N-1)}) +\mu (i,\ldots ,i)=0\cdot \] Taking the real part
of this equality, we obtain: \(\lambda =0\). So \((\Gamma
_0,\ldots ,\Gamma _{N-1})=-\mu (i,\ldots ,i)\). So the \(\Gamma
_k\) are equal. If \(N\) is odd, we have:
\[(\Gamma _0'',\ldots ,\Gamma _{N-1}'')
=(\Gamma _0,\ldots ,\Gamma _{N-1})+\lambda (1,\rho ^{-1},\ldots
,\rho ^{-(N-1)}) +\mu (i,\ldots ,i)\]
\[=\nu (1,\rho ^{(N-1)/2},\rho ^{N-1},\ldots ,\rho ^{(N-1)^2/2})\cdot \]
Taking the real part of this equality, we obtain:
\[\left (
\begin{array}{cccc}
1 & 1 & 1 & 1\\
\rho ^{-1} & \rho & \rho ^{(N-1)/2} & \rho ^{-(N-1)/2}\\
\vdots & \vdots & \vdots & \vdots \\
\rho ^{-(N-1)} & \rho ^{N-1} & \rho ^{(N-1)^2/2} & \rho
^{-(N-1)^2/2}
\end{array}
\right ) \left (
\begin{array}{c}
\lambda \\
\bar{\lambda } \\
-\nu \\
-\bar{\nu }
\end{array}
\right ) =0\cdot \] As
\[\det \left (
\begin{array}{cccc}
1 & 1 & 1 & 1\\
\rho ^{-1} & \rho & \rho ^{(N-1)/2} & \rho ^{-(N-1)/2}\\
\rho ^{-2} & \rho ^2 & \rho ^{N-1} & \rho ^{-(N-1)}\\
\rho ^{-3} & \rho ^3 & \rho ^{3(N-1)/2} & \rho ^{-3(N-1)/2}
\end{array}
\right )\]
\[=(\rho-\rho ^{-1})
(\rho ^{(N-1)/2}-\rho ^{-1}) (\rho ^{-(N-1)/2}-\rho ^{-1}) (\rho
^{(N-1)/2}-\rho) (\rho ^{-(N-1)/2}-\rho) (\rho ^{-(N-1)/2}-\rho
^{(N-1)/2})\ne 0,\] we have: \(\lambda =\nu =0\). Again, the
\(\Gamma _k\) are equal. Thus, we have just showed that for a
relative equilibrium with \(N\ge 4\) vortices at the vertices of a
regular polygon, the vorticities are always equal.\\

In fact, if from the beginning we had assumed
the total vorticity not to vanish and the center of
vorticity to be the geometric center of the polygon, this result
could have been proved in a simpler way. In this case,
for a relative equilibrium, the motion
is a {\it choreography}: the \(N\) vortices chase each other on
the same curve with the same phase shift between two vortices. Now
it can be shown that this cannot occur for distinct vorticities
(Celli, 2003).\\

Let us now consider the more general problem obtained when we add
a vortex at the geometric center of our polygon of vortices. It is
obvious that if this new configuration is a relative equilibrium,
then the previous configuration formed of a polygon is also a
relative equilibrium. Thus, for a regular polygon with \(N\ge 4\)
vortices and a vortex at the center in relative equilibrium, the
vorticities on the polygon have to be equal according to the
previous result. Finally, let us study
the case \(N=3\): for which values of the vorticities is an
equilateral
triangle of vortices with a vortex at its center in relative
equilibrium? The velocities \(v_0\), \(v_1\), \(v_2\) of the
vortices with vorticities \(\Gamma _0\), \(\Gamma _1\), \(\Gamma
_2\) at the vertices \(z_0=1\), \(z_1=\rho \), \(z_2=\rho ^2\) of
the triangle have the following expression:
\[v_k=\frac{i}{3}\sum _{l\ne k} \Gamma _l(z_k-z_l)+i\Gamma z_k
=\frac{i}{3}((\Gamma _0+\Gamma _1+\Gamma _2+3\Gamma )z_k -(\Gamma
_0z_0+\Gamma _1z_1+\Gamma _2z_2)),\] where \(\Gamma \) is the
vorticity at the center. The velocity of the vortex at the center
has the following expression:
\[v=-i(\Gamma _0z_0+\Gamma _1z_1+\Gamma _2z_2)\cdot \]
Thus, we have:
\[\frac{v_k-v}{z_k}
=\frac{i}{3}((\Gamma _0+\Gamma _1+\Gamma _2+3\Gamma )
+\frac{2}{z_k}(\Gamma _0z_0+\Gamma _1z_1+\Gamma _2z_2))\]
\[=\frac{i}{3}((\Gamma _0+\Gamma _1+\Gamma _2+3\Gamma )
+\frac{2}{z_k}(\Gamma _0+\rho \Gamma _1+\rho ^2\Gamma _2)) \cdot
\] This quantity has to be independent of \(k\), which is
equivalent to the condition:
\[\Gamma _0+\rho \Gamma _1+\rho
^2\Gamma _2=(\Gamma _0-\Gamma _2)+(\Gamma _1-\Gamma _2)\rho =0\cdot \]
So \(\Gamma _0=\Gamma _1=\Gamma _2\).

\vskip1cm

\renewcommand{\thesection}{\Roman{section}}
\section{Two polygons with the same vorticity on each polygon}
\renewcommand{\thesection}{\arabic{section}}

\begin{figure}[h]
  \centering
     \includegraphics[scale=.7]{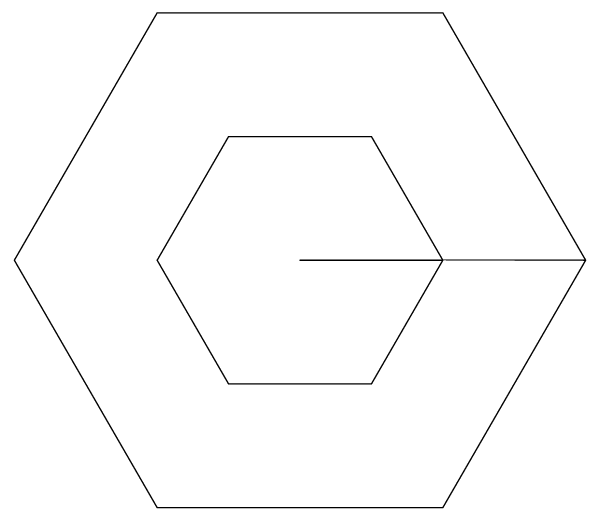}
     \includegraphics[scale=.7]{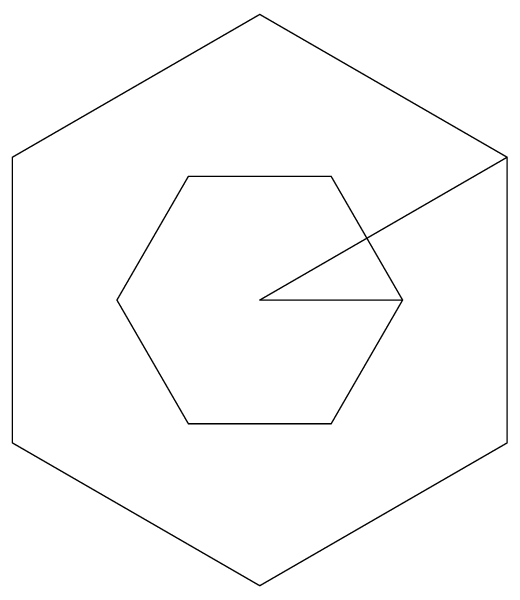}
       \caption{Symmetric configuration (on the left)
       and staggered configuration (on the right).}
       \label{symmetry}
       \end{figure}

Let us consider two concentric regular \(n\)-gons (\(n\ge 2\)),
with a vortex at each vertex, in relative equilibrium. We assume
that the vortices of a same polygon have the same non-zero
vorticity \(\Gamma _1\) or \(\Gamma _2\). The configuration of the
\(N=2n\) vortices cannot be rigidly translating (with non-zero
velocity) as, by a symmetry argument, the velocity would have to
vanish. By symmetry again, the common center of the two polygons
has zero velocity. So the motion is either a rotation around the
center of the polygons (with non-zero angular velocity) or an
absolute equilibrium. Let us remark that if we had first assumed
the center of the polygons to have zero velocity, then we would
not have needed to assume the two polygons to have the same number
of vertices (Aref et al., 2002). It can also be shown that the
angle between the two polygons has to be equal to \(0\) (the
symmetric case) or \(\pi /n\) (the staggered case),
as in figure \ref{symmetry} (Aref et al., 2002).\\

The relative equilibria formed of two concentric polygons, with
the same vorticity on each polygon, are studied in Havelock, 1931
(where the case \(\Gamma _1=-\Gamma _2\) is stressed) and Aref et
al., 2002 (where the solution of the case \(\Gamma _1=\Gamma _2 \)
can be found), among others. The simpler relative equilibria
formed of a polygon and a particle with zero vorticity can easily
be computed (see Morton, 1933; Aref et al., 2002). The more
general problem with three polygons was also solved in the
particular case where the vorticities of the three polygons are
equal (Aref and van Buren, 2005). We also know that, for three polygons,
the number of
relative equilibria corresponding to a generic set of vorticities
(identical on a same polygon) is finite (O'Neil, 2007). The
purpose of this section is to provide an exhaustive classification
of all the relative equilibria formed of two polygons,
which will be valid for every \(\Gamma _1\) and \(\Gamma _2\).\\

\begin{figure}[h]
  \centering
     \includegraphics{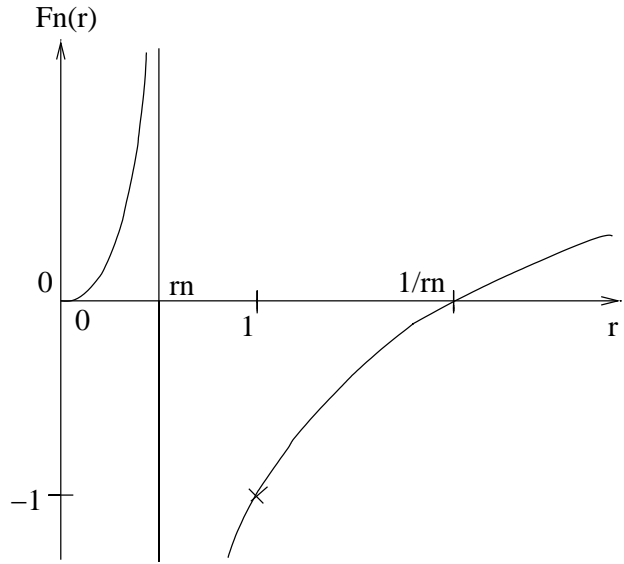}
       \caption{The function \(F_n\).}
       \label{fn}
       \end{figure}

In the symmetric case, it can be shown (see Aref et al., 2002)
that the configuration is a relative equilibrium when it satisfies
the following condition:
\[r^{n+2}-\left ( \frac{2n}{n-1}+\gamma \right )r^n
-\left ( 1+\frac{2n}{n-1}\gamma \right ) r^2+\gamma =0,\]
where \(r\) is the ratio of the radii of the two \(n\)-gons,
and \(\gamma \) is the ratio of the vorticities.
This equation remains invariant if we
change \(r\) by \(1/r\) and \(\gamma \) by
\(1/\gamma \), which corresponds to
exchanging the two polygons.
In order to solve it, it seems relevant to
separate the variables \(r\) and \(\gamma \):
\[r^{n+2}-\frac{2n}{n-1}r^n-r^2
=\left (r^n+\frac{2n}{n-1}r^2-1\right ) \gamma \cdot \]
It is easy to see that the factor \(r^n+\frac{2n}{n-1}r^2-1\)
vanishes for only one value of \(r\), which is less than \(1\),
that we denote by \(r_n\).
We then have:
\[r_n^{n+2}-\frac{2n}{n-1}r_n^n-r_n^2
=r_n^2\left (
\left (r_n^n+\frac{2n}{n-1}r_n^2-1\right )
-\frac{2n}{n-1}r_n^{n-2}-\frac{2n}{n-1}r_n^2\right )\]
\[=-\frac{2n}{n-1}r_n^2\left (r_n^{n-2}+r_n^2\right )<0\cdot \]
So in this case, the configuration is a symmetric relative equilibrium
for no value of \(\gamma \). For any other value of \(r\),
we obtain the following relation:
\[\gamma =\frac{r^{n+2}-\frac{2n}{n-1}r^n-r^2}
{r^n+\frac{2n}{n-1}r^2-1}=F_n(r)\cdot \]
Taking its derivative, we
get:
\[F_n'(r)=\frac{2r\left (r^{2n}+1+\frac{n^2}{n-1}r^{n-2}
\left ((r^2-1)^2+\frac{2(2n-1)}{n^2(n-1)}r^2\right ) \right
)}{\left (r^n+\frac{2n}{n-1}r^2-1\right )^2}>0\cdot \] We thus can
plot the graph of \(F_n\) (figure \ref{fn}). Its intersections
with the horizontal lines \(\gamma =\mbox{constant}\) provide the
relative equilibria corres\-pon\-ding to a given value of the
vorticities. The intersections with the vertical lines
\(r=\mbox{constant}\) provide the relative equilibria
corres\-pon\-ding to a given value of the radii.\\

Then we can see that for a given value of the vorticities
\(\Gamma _1\) and \(\Gamma _2\):\\
- If \(\Gamma _1\) and \(\Gamma _2\) have the same sign and
\(\Gamma _1\ne \Gamma _2\), there are exactly two symmetric
relative equilibria (up to similarities). For one of them, the
vortices with the larger vorticity (in absolute value) are on the
larger polygon. For the other, they are on the smaller polygon.
The same result had been obtained for the analogous problem in
celestial mechanics, where the masses (quantities analogous to the
vorticities) are always positive, so they have the same sign
(Moeckel and Sim\'o, 1995). For our two relative equilibria, the
ratio of the small radius
to the large radius is less than \(r_n\).\\
- If \(\Gamma _1=\Gamma _2\), there is exactly one symmetric
relative equilibrium. This is a classical result (see Aref et al.,
2002). Here, by symmetry, each solution \(r\) of the previous case
is the multiplicative inverse of the other solution. So these two
solutions define the same
relative equilibrium, up to an exchange of the polygons.\\
- If \(\Gamma _1\) and \(\Gamma _2\) have
opposite signs and \(|\Gamma _1|\ne |\Gamma _2|\),
there is exactly one symmetric relative equilibrium.
The vortices with the larger vorticity (in absolute value)
are on the smaller polygon. The ratio of the small radius
to the large radius is greater than \(r_n\).\\
- If \(\Gamma _1=-\Gamma _2\), there is no symmetric
relative equilibrium, as the unique solution \(r\) is \(1\):
the polygons would be coincident. This is a classical result
(see Aref, 1982).\\

Moreover, for a given configuration:\\
- If the ratio of the small radius
to the large radius is equal to \(r_n\),
the configuration is a symmetric relative equilibrium
for no value of \(\Gamma _1\) and \(\Gamma _2\).
Otherwise there exist exactly one \(\Gamma _1\)
and one \(\Gamma _2\) (up to a coefficient of proportionality)
such that the configuration is a symmetric
relative equilibrium.\\
- If this ratio is less than \(r_n\), the vorticities
have the same sign.\\
- If this ratio is greater than \(r_n\), the vorticities
have opposite signs and \(|\Gamma _1|\ne |\Gamma _2|\).
The vortices with the larger vorticity
(in absolute value) are on the smaller polygon.\\

In the staggered case, it can be shown (see Aref et al., 2002)
that the configuration is a relative equilibrium when it satisfies
the following condition:
\[r^{n+2}-\left ( \frac{2n}{n-1}+\gamma \right )r^n
+\left ( 1+\frac{2n}{n-1}\gamma \right ) r^2-\gamma =0\cdot \]
This equation remains invariant if we
change \(r\) by \(1/r\) and \(\gamma \) by
\(1/\gamma \), which corresponds to
exchanging the two polygons.
It is equivalent to:
\[r^{n+2}-\frac{2n}{n-1}r^n+r^2
=\left (r^n-\frac{2n}{n-1}r^2+1\right ) \gamma \cdot \]

Let us first consider the amazing case \(n=4\).
Then this equation becomes equivalent to:
\[\left (r^4-\frac{8}{3}r^2+1\right )(r^2-\gamma )=0\cdot \]
Thus, the staggered configuration corresponding to
\[r=r_4'=\sqrt{\frac{4-\sqrt{7}}{3}}\mbox{ or }
r=r_4''=\sqrt{\frac{4+\sqrt{7}}{3}}=\frac{1}{r_4'}\]
is a relative equilibrium {\it for any vorticities}.
For vorticities \(\Gamma _1\) and \(\Gamma _2\)
with the same sign, there is exactly one more relative equilibrium
(up to similarities),
corresponding to \(r=\sqrt{\Gamma _2/\Gamma _1}\).
For vorticities \(\Gamma _1\) and \(\Gamma _2\)
with opposite signs, there is no other relative equilibrium.
Every staggered configuration with \(r\ne r_4'\), \(r_4''\),
is a relative equilibrium for exactly one \(\Gamma _1\)
and one \(\Gamma _2\) (up to a coefficient of proportionality),
which have the same sign.\\
There exist exactly one value \(\tilde{\Gamma} _1\) of \(\Gamma
_1\) and one value \(\tilde{\Gamma} _2\) of \(\Gamma _2\) (up to a
coefficient of proportionality) such that the staggered relative
equilibrium corresponding to \(r=r_4'\), for instance, is in fact
an absolute equilibrium. This is due to the linearity of the
angular velocity \(\omega \), seen as a function of \(\Gamma _1\),
\(\Gamma _2\). It has rank \(1\), because when \(\Gamma _2\) tends
to \(0\), the angular velocity of polygon \(1\) is different from
\(0\). So its kernel has dimension \(2-1=1\). This absolute
equilibrium allows to construct relative equilibria which are
solutions of Helmholtz's equations for several values of the
vorticities: observing the motion of the vortices would not be
enough to determine the vorticities! The relative equilibria
formed by a regular polygon with equal vorticities and a vortex at
its center are known examples of such solutions. In the present
case, as \(\omega (\Gamma _1,\Gamma _2)\) is linear, any
vorticities \((\Gamma _1, \Gamma _2)\) will define the same
rotation as the vorticities \((\Gamma _1+\lambda \tilde{\Gamma}
_1, \Gamma _2+\lambda \tilde{\Gamma} _2)\) for every coefficient
\(\lambda \), as:
\[\omega (\Gamma _1+\lambda \tilde{\Gamma} _1,
\Gamma _2+\lambda \tilde{\Gamma} _2)
=\omega (\Gamma _1,\Gamma _2)
+\lambda \omega (\tilde{\Gamma} _1,\tilde{\Gamma} _2)
=\omega (\Gamma _1,\Gamma _2)\cdot \]

\begin{figure}[h]
  \centering
     \includegraphics{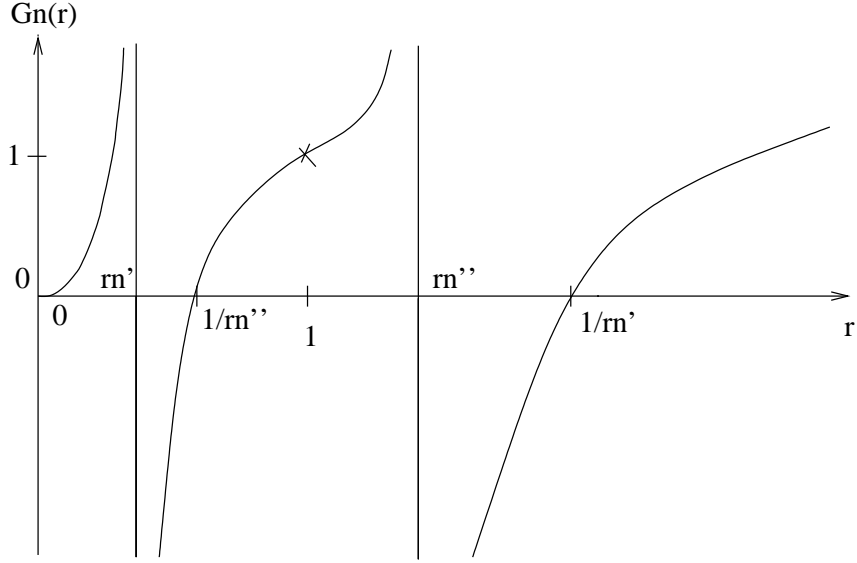}
       \caption{The function \(G_n\), \(n\ge 5\).}
       \label{gn}
       \end{figure}

From now on, we assume that \(n\ne 4\).
If \(n=2\), the factor \(r^n-\frac{2n}{n-1}r^2+1\)
only vanishes for \(r=r_2'=1/\sqrt{3}\).
If \(n=3\) or \(n\ge 5\), this factor has a unique critical point,
and it is positive at \(0\) and \(+\infty \)
and negative at \(1\). So it vanishes exactly
for two values \(r_n'<1<r_n''\) of \(r\).
In every case, we have, for \(r=r_n'\) or \(r_n''\):
\[r^{n+2}-\frac{2n}{n-1}r^n+r^2
=r^2\left (
\left (r^n-\frac{2n}{n-1}r^2+1\right )
-\frac{2n}{n-1}r^{n-2}+\frac{2n}{n-1}r^2\right )\]
\[=-\frac{2n}{n-1}r^2\left (r^{n-2}-r^2\right )\ne 0\]
as \(n\ne 4\) and \(r_n'\), \(r_n''\ne 1\).
So for \(r=r_n'\) or \(r_n''\), the configuration is a
staggered relative equilibrium for no value of \(\gamma \).
For any other value of \(r\), we obtain the following relation:
\[\gamma =\frac{r^{n+2}-\frac{2n}{n-1}r^n+r^2}
{r^n-\frac{2n}{n-1}r^2+1}=G_n(r)\cdot \]
Taking its derivative, we
get:
\[G_n'(r)=2r^{n+1}\frac{
(r^n+r^{-n})
-\frac{n^2}{n-1}(r^2+r^{-2})
+\frac{2(n^3-n^2-2n+1)}{(n-1)^2}
}{\left (r^n-\frac{2n}{n-1}r^2+1\right )^2}\]
\[=\frac{4r^{n+1}H_n(\ln (r))}
{\left (r^n-\frac{2n}{n-1}r^2+1\right )^2},\]
\[\mbox{where } H_n(u)=
\cosh (nu)-\frac{n^2}{n-1}\cosh (2u)+\frac{n^3-n^2-2n+1}{(n-1)^2}
\cdot \]
If \(n\ge 5\), we have:
\[H_n''(u)=
n^2\cosh (nu)-\frac{4n^2}{n-1}\cosh (2u)\ge 0,\]
as
\[\left \{
\begin{array}{c}
n^2\ge \frac{4n^2}{n-1}>0\\
\cosh (nu)\ge \cosh (2u)>0
\end{array}
\right . \]
This allows to prove that \(H_n'(u)\) has the same sign
as \(u\), so that \(H_n\) has a minimum at \(0\) and:
\[H_n(u)\ge H_n(0)=\frac{n(n-4)+2}{(n-1)^2}>0\cdot \]
So we have: \(G_n'(r)>0\). We thus can plot the graph
of \(G_n\) for \(n\ge 5\) (figure \ref{gn}).\\

Then we can see that for a given value of the vorticities
\(\Gamma _1\) and \(\Gamma _2\):\\
- If \(\Gamma _1\) and \(\Gamma _2\) have the same sign and
\(\Gamma _1\ne \Gamma _2\), there are exactly three staggered
relative equilibria. For one of them, the ratio of the small
radius to the large radius is greater than \(1/r_n''\) (so it is
greater than \(r_n'\)), and the vortices with the larger vorticity
(in absolute value) are on the larger polygon. For the two others,
the ratio of the small radius to the large radius is less than
\(r_n'\) (so it is less than \(1/r_n''\)). For one of these two
relative equilibria, the vortices with the larger vorticity are on
the larger polygon. For the other, they are on the smaller polygon.\\
- If \(\Gamma _1=\Gamma _2\), there are exactly two staggered
relative equilibria. This is a classical result (see Aref et al.,
2002). One of them corresponds to the regular \(2n\)-gon. For the
other, the ratio of the small radius to the large radius is less
than \(r_n'\) (so it is less than \(1/r_n''\)). It corresponds to
the two last relative equilibria of the previous case. Here, by
symmetry, each value of \(r\) for these two solutions is the
multiplicative inverse of the other value. So these two solutions
define the same relative
equilibrium, up to an exchange of the polygons.\\
- If \(\Gamma _1\) and \(\Gamma _2\) have opposite signs and
\(|\Gamma _1|\ne |\Gamma _2|\), there are exactly two staggered
relative equilibria. For one of them, the vortices with the larger
vorticity are on the larger polygon. For the other, they are on
the smaller polygon. For both, the ratio of the small radius to
the large radius is between \(r_n'\) and \(1/r_n''\).\\
- If \(\Gamma _1=-\Gamma _2\), there is exactly one staggered
relative equilibrium. This is a classical result (see Aref, 1982).
Here, by symmetry, each solution \(r\) of
the previous case is the multiplicative inverse of the other
solution. So these two solutions define the same relative
equilibrium, up to an exchange of the polygons.\\

Moreover, for a given configuration:\\
- If the ratio of the small radius to the large radius is equal to
\(r_n'\) or \(1/r_n''\), the configuration is a staggered relative
equilibrium for no value of \(\Gamma _1\) and \(\Gamma _2\).
Otherwise there exist exactly one \(\Gamma _1\) and one \(\Gamma
_2\) (up to a coefficient of proportionality) such that the
configuration is a staggered relative equilibrium.\\
- If this ratio is less than \(r_n'\) or greater than \(1/r_n''\),
the vorticities have the same sign. When it is greater than
\(1/r_n''\), the vortices with the larger vorticity are on the
larger polygon. When it is equal to \(1\) (then the configuration
is a regular \(2n\)-gon), all the vorticities are equal.\\
- If this ratio is between \(r_n'\) and \(1/r_n''\),
the vorticities have opposite signs.\\

\begin{figure}[h]
  \centering
     \includegraphics[scale=.6]{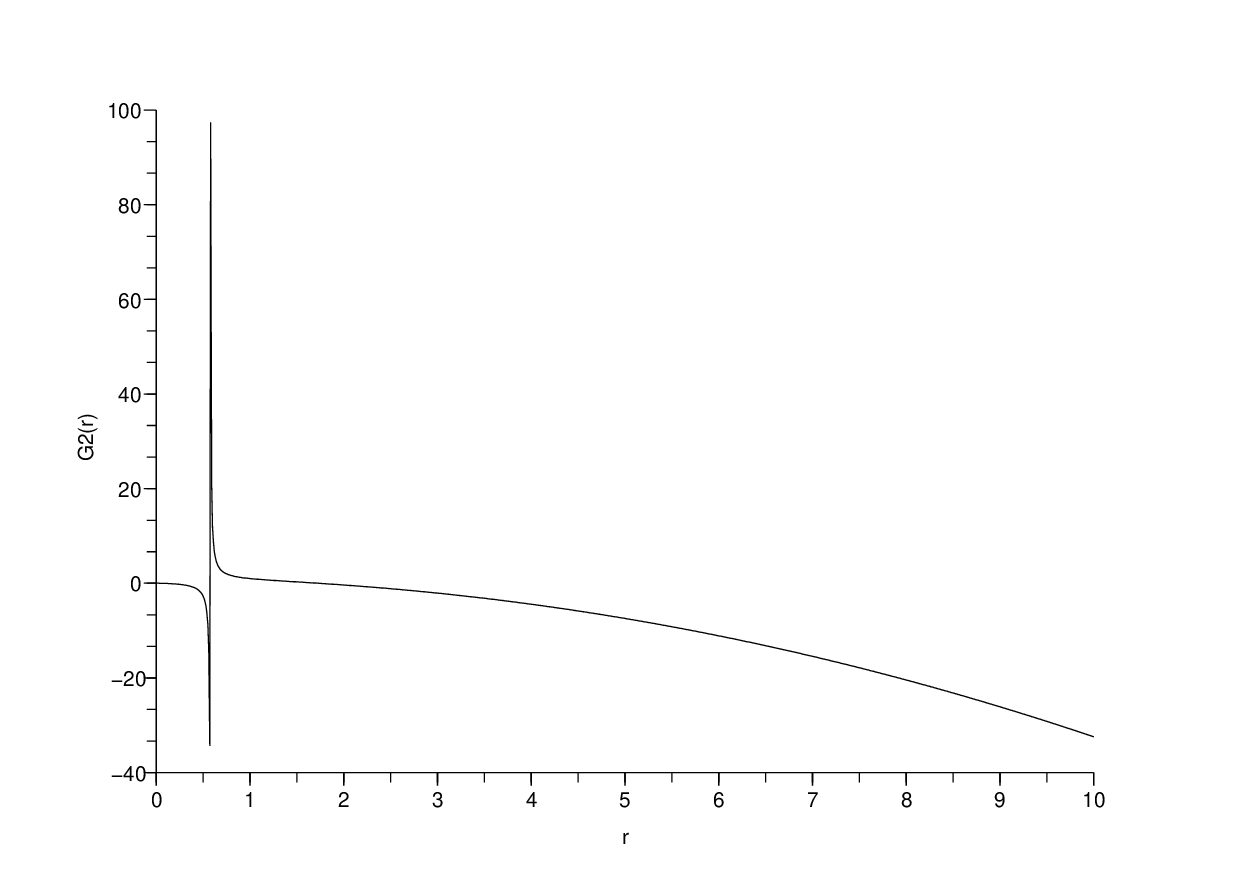}
       \caption{The function \(G_2\).}
       \label{g2}
       \end{figure}

It is easy to study the case \(n=2\) (figure \ref{g2}) by checking
that:
\[H_2(u)=1-3\cosh (u)<0\cdot \]

\begin{figure}[h]
  \centering
     \includegraphics[scale=.35]{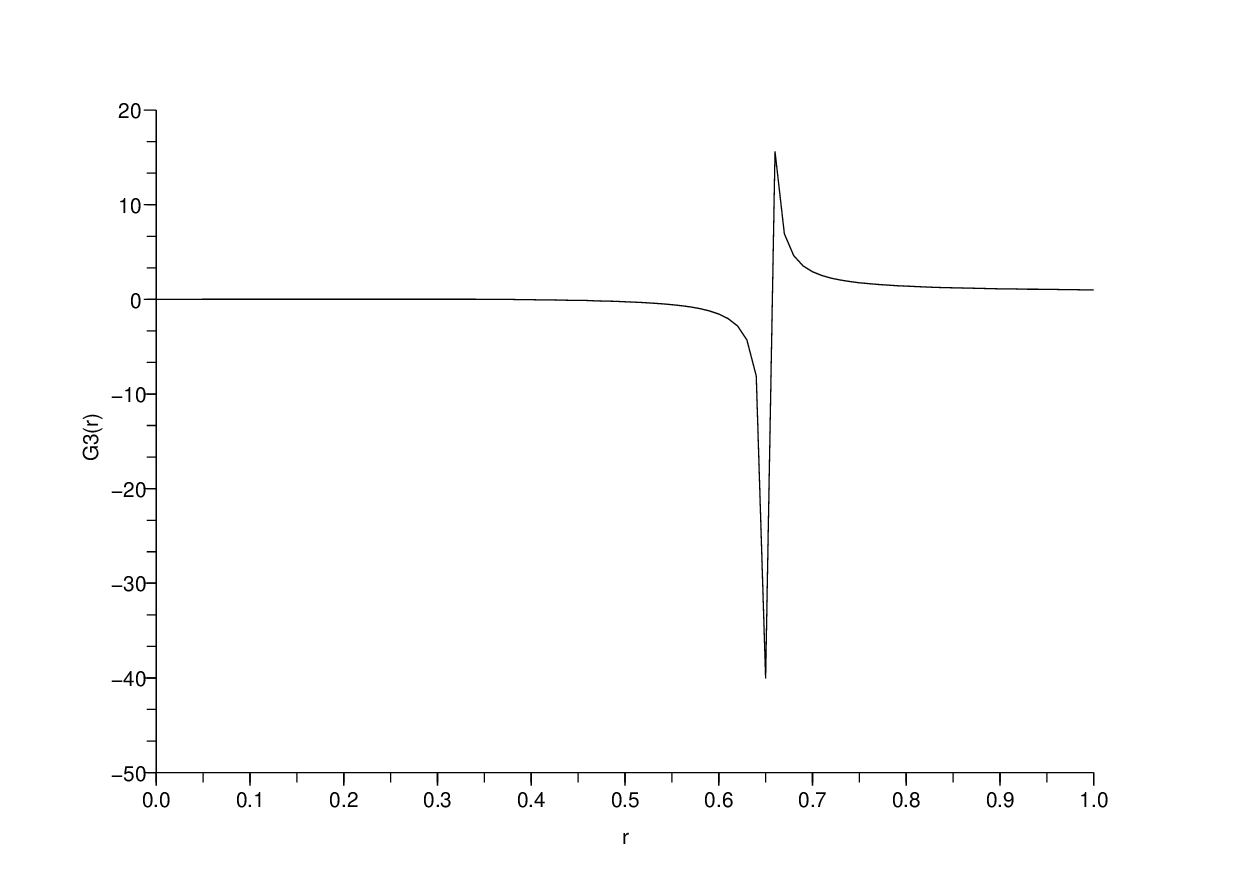}
     \includegraphics[scale=.35]{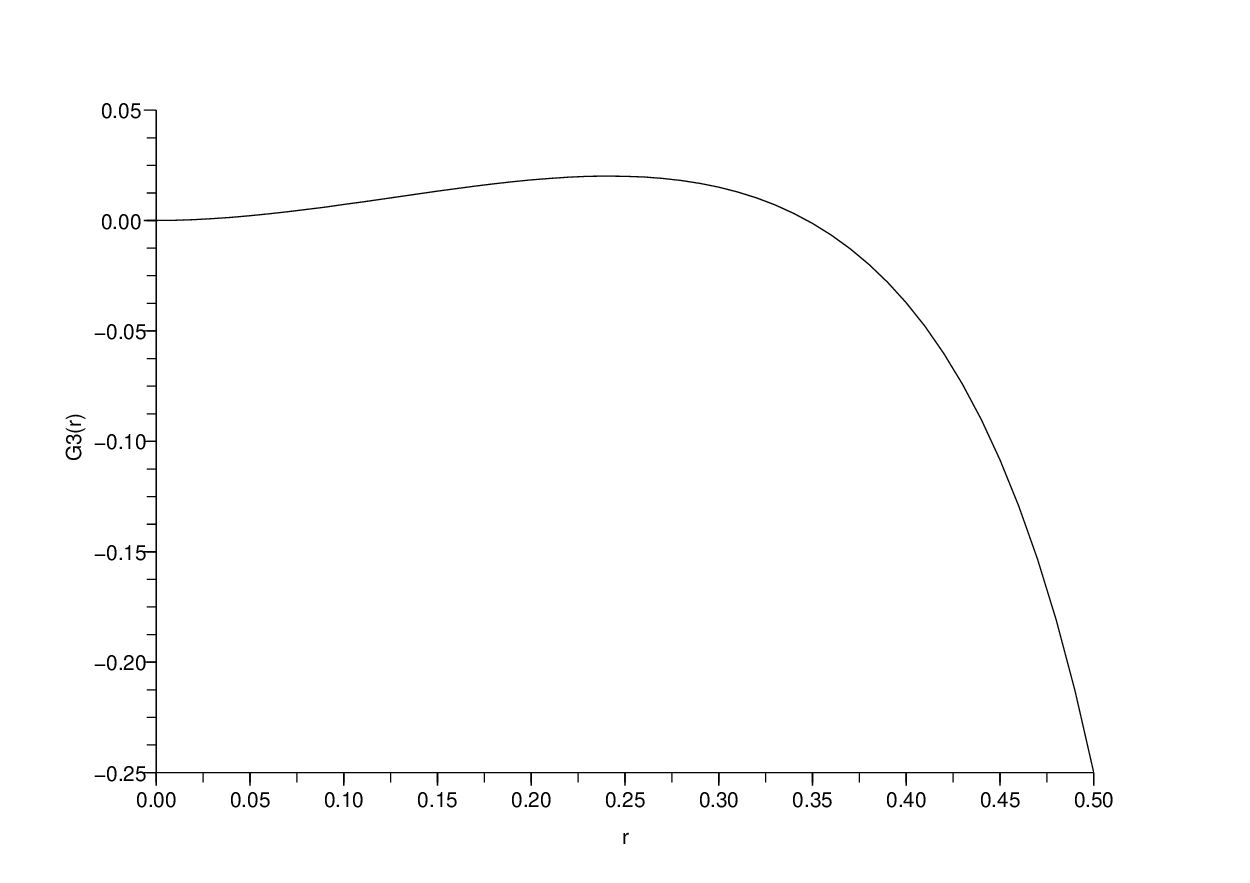}
       \caption{The function \(G_3\) for \(0\le r\le 1\) (on the left)
and for \(0\le r\le 0.5\) (on the right, where we can see the
local maximum corresponding to \(s_3'\)).}
       \label{g3}
       \end{figure}

We can study the case \(n=3\) by checking that:
\[H_3^{(4)}(u)>0\cdot \]
This allows to prove that the numerator of \(G_3'(r)\) vanishes
exactly for two values \(s_3'<s_3''\) of \(r\), each one being the
multiplicative inverse of the other. We obtain:
\[r_3'\approx 0.6527036\mbox{, }r_3''\approx 2.8793852\mbox{, }
s_3'\approx 0.2418796\mbox{, }s_3''\approx 4.1342878,\] which
allows to plot the graph of \(G_3\) (figure
\ref{g3}). For convenience, we plotted the graph of the function
\(G_3\) only for \(0\le r\le 1\), which corresponds to the case
where polygon \(2\) is inside. Then it is easy to plot the graph
for \(r\ge 1\), as changing \(r\) by \(1/r\) corresponds to
exchanging the two polygons, or to changing \(\gamma \) by
\(1/\gamma \). Thus, the \(r\ge 1\) corresponding to a given value
of \(\gamma \) are the multiplicative inverses of the \(r\le 1\)
corresponding to \(1/\gamma \) in figure \ref{g3}.\\

For all \(n\), we can determine which relative equilibria are, in
fact, absolute equilibria. A configuration formed of two
concentric regular \(n\)-gons (with center at the origin) is an
absolute equilibrium when a vortex of polygon \(1\) (at point
\(z_1\)) and a vortex of polygon \(2\) (at point \(z_2\)) have
zero velocity. Repeating the calculation made in Aref et al.
(2002), we can show that this condition is equivalent to:
\[\frac{\Gamma _2n}{1-\left (\frac{\bar{z}_2}{\bar{z}_1}\right
)^n}+\frac{\Gamma _1(n-1)}{2}=\frac{\Gamma _1n}{1-\left
(\frac{\bar{z}_1}{\bar{z}_2}\right )^n}+\frac{\Gamma
_2(n-1)}{2}=0\cdot \] This is also equivalent to:
\[\left (\frac{z_2}{z_1} \right )^n=
1+\frac{2n}{n-1}\gamma =\frac{1}{1+\frac{2n}{n-1}\frac{1}{\gamma
}}\cdot \] The identity of the two last members is equivalent to:
\[1+\frac{2n}{n-1}\gamma =-\gamma ^2\cdot \]
So the configuration is an absolute equilibrium when:
\[\left (\frac{z_2}{z_1} \right )^n
=-\gamma ^2=1+\frac{2n}{n-1}\gamma \cdot \] We obtain the value of
\(\gamma \) by solving the second degree equation which expresses
the identity of the two last members. Then the identity of the two
first members gives the value of \(z_2/z_1\). Thus, we can see
that only in the staggered case
the configuration can be an absolute equilibrium,
for the following values:
\[\gamma =-\left (
\frac{n}{n-1}+\sqrt{\left ( \frac{n}{n-1}\right)^2-1} \right
)\mbox{, } r=\left ( \frac{n}{n-1}+\sqrt{\left (
\frac{n}{n-1}\right)^2-1} \right )^{\frac{2}{n}}\] (or for the
multiplicative inverses of these values, obtained after exchanging the two polygons).\\

When \(n\) tends to \(+\infty \), we obtain the relative
equilibria formed of two concentric homogeneous circles. The
quantities \(\Gamma _1\) and \(\Gamma _2\) now are the linear
vorticity densities of the circles. The configuration can equally
be seen as a symmetric or staggered configuration, and we have:
\[F_{\infty }(r)=G_{\infty }(r)= \left \{
\begin{array}{c}
\frac{1}{\frac{1}{r^2}-2}\mbox{ if } r<1\\
r^2-2\mbox{ if } r>1
\end{array} \right .\]
This function has the same variations as the functions \(F_n\)
plotted in figure \ref{fn}, with \(r_{\infty}=1/\sqrt{2}\), and
the functions \(G_n\) plotted in figure \ref{gn}, with
\(r_{\infty}'=1/\sqrt{2}\) and \(r_{\infty}''=1\). Thus, the
results stated above, related to the symmetric and staggered
relative equilibria for finite \(n\), remain valid in the case of
two concentric homogeneous circles. There is no absolute
equilibrium as, when \(n\) tends to \(+\infty \) in the equations
of the previous paragraph, we obtain: \(r=1\).

\vskip1cm

\section*{Acknowledgements}
This research has been partially supported by CONACYT-M\'exico,
grant 128790.

\vskip1cm

\section*{References}

\(\)

Aref, H., "Point vortex motions with a center of symmetry", Phys.
Fluids \textbf{25}, no. 12, 2183-2187 (1982).\\

Aref, H., "Point vortex dynamics: a classical mathematics
playground", J. Math. Phys. \textbf{48},
65401.1-65401.23 (2007).\\

Aref, H., Newton, P. K., Stremler, M. A., Tokieda, T., Vainchtein,
D. L., "Vortex crystals", Adv. Appl. Mech. \textbf{39}, 1-79 (2002).\\

Aref, H., van Buren, M., "Vortex triple rings",
Phys. Fluids \textbf{17}, 57104.1-57104.21 (2005).\\

Celli, M., "Sur les distances mutuelles d'une chor\'egraphie \`a
masses distinctes", Comptes Rendus Math. Ser. I \textbf{337},
715-720 (2003).\\

Davis, C., Wang, W., Chen, S. S., Chen, Y., Corbosiero, K.,
DeMaria, M., Dudhia, J., Holland, G., Klemp, J., Michalakes, J.,
Reeves, H., Rotunno, R., Snyder, C., Xiao, Q., "Prediction of
landfalling hurricanes with the advanced hurricane
WRF model", Mon. Wea. Rev. \textbf{136}, 1990-2005 (2008).\\

Elmabsout, B., "Sur l'existence de certaines configurations
d'\'equilibre relatif dans le probl\`eme des \(N\) corps",
Cel. Mech. Dyn. Astr. \textbf{41}, 131-151 (1988).\\

Hampton, M., Moeckel, R., "Finiteness of stationary configurations
of the four-vortex problem", Trans. Amer. Math. Soc. \textbf{361},
no. 3, 1317-1332 (2009).\\

Havelock, T. H., "The stability of motion of rectilinear vortices
in ring formation", Philos. Mag. (7) \textbf{11}, 617-633 (1931).\\

Helmholtz, H., "On integrals of the hydrodynamical equations which
express vortex motion", Philos. Mag. \textbf{33},
485-512 (1858).\\

Kossin, J. P., Schubert, W. H., "Mesovortices, polygonal flow
patterns, and rapid pressure falls in hurricane-like vortices", J.
Atmos. Sci. \textbf{58}, 2196-2209 (2001).\\

Kossin, J. P., Schubert, W. H., "Mesovortices in Hurricane
Isabel", Bull. Amer. Met. Soc. \textbf{85}, issue 2, 151-153 (2004).\\

Moeckel, R., Sim\'o, C., "Bifurcation of spatial central
configurations from planar ones", SIAM J. Math. Anal. \textbf{26},
978-998 (1995).\\

Morton, W. B., "On some permanent arrangements of parallel
vortices and their points of relative rest", Proc. R. Ir. Acad. A \textbf{41},
94-101 (1933).\\

McNoldy, B. D., "Triple eyewall in Hurricane Juliette", Bull.
Amer. Met. Soc. \textbf{85}, issue 11, 1663-1666 (2004).\\

Newton, P. K., {\it The $N$-vortex problem: analytical techniques}
(Springer Verlag, New York, 2001).\\

O'Neil, K. A., "Relative equilibrium and collapse configurations
of heterogeneous vortex triple rings", Physica D \textbf{236},
123-130 (2007).\\

Perko, L., Walter, E., "Regular polygon solutions of the $N$-body
problem", Proc. Amer. Math. Soc. \textbf{94}, 301-309 (1985).\\

Yarmchuk, E. J., Gordon, M. J. V., Packard, R. E., "Observation of
stationary vortex arrays in rotating superfluid helium", Phys.
Rev. Letters \textbf{43}, no. 3, 214-217 (1979).

\end{document}